\documentclass[preprintnumbers,superscriptaddress,landscape,nofootinbib]{revtex4}
\usepackage[utf8]{inputenc}
\usepackage{amsmath}
\usepackage{amsthm}
\usepackage{amscd}
\usepackage{amssymb}
\usepackage{float}
\usepackage{braket}
\usepackage{graphicx}
\usepackage{url}
\usepackage[colorlinks=true, pdfstartview=FitV, linkcolor=red, citecolor=blue, urlcolor=blue]{hyperref}
\usepackage{slashed}
\usepackage[normalem]{ulem}

\newtheorem{theorem}{Theorem}[section]

\newtheorem{claim}[theorem]{Claim}

\theoremstyle{definition}

\theoremstyle{remark}

\begin{document}

\title{Topological Aspects of Matters and Langlands Program} 
\author{Kazuki Ikeda}
\email[]{kazuki7131@gmail.com}
\affiliation{Department of Mathematics and Statistics
$\&$ Centre for Quantum Topology and Its Applications (quanTA), University of Saskatchewan, Saskatoon, Saskatchewan S7N 5E6, Canada}

\bibliographystyle{unsrt}

\begin{abstract}
The Langlands program is a vast mathematical projection linking number theory and geometry. In high-energy physics, a connection with mirror symmetry has been suggested in string theory, but it has been little studied in low-energy physics. In the framework of the Langlands program, we present a unified description of the integer and fractional quantum Hall effect and the duality found in the fractal nature of the energy spectrum of two-dimensional block electrons, statistical physics, and quantum computation. The new unified view of existing dualism presented in this paper raises the entirely new question of how each theory of physics is connected as a piece of the Langlands program.
\end{abstract}

\maketitle

\section{Introduction}
Duality is ubiquitous in modern physics. It provides a succinct description of key connections among apparently different phenomena and consequently offers new perspectives to examine each of them. They are often referred to electric/magnetic duality, strong/weak duality or high/low duality. These dualities are closely linked in a "duality web", yet formal modeling has been illusive, thereby it will be natural to seek for an underling theoretical framework. Langlands program \cite{10.1007/BFb0079065} is a mathematical coherent conjecture that offers a compelling and captivating story to connect a wide variety of mathematical concepts in terms of duality \cite{gaitsgory2004vanishing,Frenkel:1995zp,Laumon,Hausel:2002ap,2000math.....12255F}. It is also related to high energy physics \cite{Kapustin:2006pk,gukov2008,aganagic2018quantum}. We aim at applying it to topological physics, which leads to enhance our knowledge on the "duality web". Topological invariance is a central concept in modern physics. Especially the quantum Hall effect \cite{doi:10.1143/JPSJ.39.279,PhysRevB.23.5632,0022-3719-14-23-022,PhysRevLett.45.494} has led to a number of both theoretical and experimental studies of the problem. In this article we exploits the Langlands program to offer an unified description of 2d integer and fractional quantum Hall effect (IQHE and FQHE). Consequently it elicits a part of uncanny insights into various cardinal principles of physics and also sheds light on useful applications of topological phases of matter (e.g. topological quantum computation). Traditionally the IQHE $\sigma_{xy}=\frac{e^2}{h}n~(n\in\mathbb{Z})$ has been explained by the Kubo formula \cite{doi:10.1143/JPSJ.12.570,1982PhLA...90..474W,0022-3719-15-22-005,PhysRevLett.49.405} and the Anderson localization model \cite{PhysRev.109.1492,PRUISKEN1984277}. The FQHE has been explained by the Laughlin theory \cite{PhysRevLett.50.1395} and composite particle models \cite{doi:10.1142/S0217979292000037,PhysRevLett.63.199,PhysRevLett.62.86}.

In the previous work~\cite{doi:10.1063/1.4998635} of the author, the fractal energy spectrum of the quantum Hall system, called Hofstadter butterfly, was connected to the Langlands duality of the quantum group that generates the Hamiltonian. In another work~\cite{IKEDA2018136} of the author, the quantum Hall conductance were suggested to be related to the Hecke operator and the Hecke eigensheaf. The (integer/fractional) quantum Hall effect is not only a fundamental example of a topological insulator, but is also very important in mathematical physics as a model of Chern-Simons theory, Conformal Field Theory (CFT) and integrable systems. Furthermore, it recently has important applications to quantum computation. Interestingly, duality exists in Chern-Simons theory, CFT, and quantum computation, all of which have counterparts in high-energy physics. The most famous example is the mirror symmetry, and as shown by Kapustin and Witten, they have a close relationship with the Langlands program~\cite{Kapustin:2006pk}. Investigating the relationship between the Langlands program and low-energy physics is a question that naturally follows from these facts. However, to the best of author's knowledge, there is little research on low-energy physics and the Langlands program. Therefore, as a first step in approaching this question, this paper investigates what types of duality in low-energy physics are relevant to the Langlands program. 

Surprisingly, it becomes clear that a fairly broad range of physics is associated with the Langlands program, including not only the physics of topological insulators, but also statistical physics and quantum computation. In fact, there is always an algebraic geometrical background behind topological phenomena (topological charges, topological layers, topological currents) and theories based on topological methods (topological field theories), which have been actively studied in recent years. Therefore, it is worth investigating low-energy physics from the perspective of algebraic geometry, which is expected to allow the theory to be developed from a unified viewpoint. In particular, in terms of the theory of physical properties, the Brilliant zone is a Riemann surface, and energy bands are formulated as a theory of vector bundles on Riemann surfaces. Since these theories are the best known in algebraic geometry, it is natural to pursue an algebraic geometry approach to the theory of condesed matter physics. Even statistical mechanics, seemingly unrelated to algebraic geometry, has deep connections to band theory. Furthermore, in quantum computers, adiabatic quantum computation and quantum annealing can provide the same computational power as topological quantum computation, which is based on the fractional quantum Hall effect. Therefore, it is quite natural to attempt to clarify the relationship between low-energy physical theory, including statistical theory and quantum computation, and the Langlands program.

This work is organized as follows.

\section{Geometric Langlands and Integer Quantum Hall Effect}
Let $X$ be a complex curve (Riemann surface) and $G$ be a Lie group. The geometric Langlands correspondence is an equivalence between the derived category $D(\text{Bun}_G(X))$
of $D$-modules on the moduli stack $\text{Bun}_G(X)$ and the derived category $\text{QCoh(Loc}_{^LG} (X))$ of quasicoherent sheaves on the stack $\text{Loc}_{^LG}(X)$. For $G = GL_1$ the moduli stack $\text{Bun}_G$ is the Picard variety $\text{Pic}$ of $X$ classifying line bundles on $X$

In our case, the  Brilluoin zone $\text{Jac}(X)$ is the Jacobian of $X$ and parametrizes distinct $U(1)$-representations of $\pi_1(X)$. The degree zero component $\text{Pic}_0$ corresponds to $\text{Jac}(X)$. In general the Hall conductance is given by the sum of the Chern number of line bundles $\mathcal{L}_i\in \text{Pic}$ below the Fermi level  
\begin{equation}
    \sigma_{xy}=\sum_{i}c_1(\mathcal{L}_i). 
\end{equation}

We consider the Abel-Jacobi map $j:X\to\text{Jac}$ sending $p\in X$ to the line bundle $\mathcal{O}_X(p-p_0)$ for some fixed reference point $p_0\in X$,
\begin{equation}
j(p)=\int_{p_0}^p\omega,
\end{equation}
where $\omega$ a basis of the space of holomorphic differentials on $X$. By the Abel-Jacobi map, we have an isomorphism as complex manifolds $\mathbb{C}/\Lambda\simeq\text{Jac}$, where $\Lambda$ is the lattice spanned by the integrals of $\omega$ over the one-cycles in $X$. 

Now let us consider a holomorphic line bundle $\mathcal{L}$ with connection on $X$. A local system is a pair $\mathcal{L}=(\mathcal{L},\nabla)$, where $\nabla$ is a holomorphic connection on $\mathcal{L}$. Due to the Forier-Mukai transformation, $\mathcal{L}=(\mathcal{L},\nabla)$ is mapped to a $\mathcal{D}$-module $\mathcal{F}=(\mathcal{F},\tilde{\nabla})$, where $\mathcal{F}$ is a holomorphic line bundle on Jac and $\tilde{\nabla}$ is a
flat holomorphic connectionn $\mathcal{F}$, which is called a Hecke eigensheaf with respect to $\mathcal{L}$. We can use a Berry connection for $\tilde{\nabla}$ and then the Hecke functor should be understood as the Wilson loop defined by the Berry connection. In addition we can consider more general Abel-Jacobi map $\pi_d : S^dX \to \text{Pic}_d$, sending $(p_i)_{i=1,\cdots,d}$ to the line bundle $\mathcal{O}_X(p_1+\cdots+p_{d}-dp_0)$,  and obtain the sheaves on the other components $\text{Pic}_d$ which carries non-trivial Hall conductance and satisfies the Hecke eigensheaf property.

Let $T^2_{BZ}$ be the 2d toric Brillouin Zone. We denote by $\mathbb{C}^m_{T^2_{BZ}}=T^2_{BZ}\times \mathbb{C}^m$ the constant sheaf whose fibre is $\mathbb{C}^m$. A sheaf $\mathcal{L}$ of $\mathbb{C}_{T^2_{BZ}}$-module is called a rank $m$ local system\index{local system} if every point $p\in T^2_{BZ}$ has an open neighbor $U$ on which $\mathcal{L}|_U$ is isomorphic to $\mathbb{C}^m_{T^2_{BZ}}$. For any $p\in T^2_{BZ}$, the category $\text{Loc}(T^2_{BZ})$ of local systems and the category $\text{Rep}(\pi_1(T^2_{BZ},p))$ of finite dimensional representations of $\pi_1(T^2_{BZ},p)$ are equivalent. Such a functor $\rho$ is given as follows. Let $[\gamma]\in\pi_1(T^2_{BZ},p)$ be a loop. Parallel translation of sections along $\gamma$ defines a monodromy representation $\rho(\mathcal{L},p):\pi_1(T^2_{BZ},p)\to GL_m(\mathbb{C})$. \if{Let $\text{Rep}(\pi_1(T^2_{BZ},p),m)$ be the category of $\pi_1(T^2_{BZ},p)$'s $m$-dimensional representations. }\fi A functor defined by the monodromy representation gives an equivalence of categories 
\begin{equation}
\rho: \text{Loc}(T^2_{BZ})\to \text{Rep}(\pi_1(T^2_{BZ},p))
\end{equation}
Moreover the category $\text{Conn}(T^2_{BZ})$ of integrable connections on $T^2_{BZ}$ is equivalent to $\text{Loc}(T^2_{BZ})$. Let $\mathcal{E}=(\mathcal{E},\nabla)$ be a rank $m$ integrable connection on $T^2_{BZ}$. For a given $\mathcal{E}$, horizontal sections form the subsheaf $\mathcal{E}^\nabla=\{A\in \mathcal{E}: \nabla A=0\}$. Then the functor 
\begin{align}
\begin{aligned}
E: \text{Conn}(T^2_{BZ})&\to \text{Loc}(T^2_{BZ})\\
\mathcal{E}&\mapsto \mathcal{E}^\nabla
\end{aligned}
\end{align}
yields an equivalence of categories. Let $G=GL_m(\mathbb{C})$ and $\text{Bun}_m(T^2_{BZ})$ be the moduli stack of rank $m$ bundles on $T^2_{BZ}$. The $n$-th Hecke correspondence $\mathcal{H}_n$ is the moduli space of $(\mathcal{M},\mathcal{M}', p)$, where sections of $\mathcal{M}'\in \text{Bun}_m$ are that of $\mathcal{M}\in \text{Bun}_m$ having a pole of order $n$ at $p$. \if{So $\mathcal{M}/\mathcal{M}'$ is isomorphic to the direct sum $\mathcal{O}^{\oplus n}_p$ of $n$ skyscraper sheaves.}\fi Let $h^{\rightarrow}(\mathcal{M},\mathcal{M}', p)=\mathcal{M}'$, $h^{\leftarrow}(\mathcal{M},\mathcal{M}', p)=\mathcal{M}$ and $\text{supp}(\mathcal{M},\mathcal{M}', p)=p$. The fiber $(h^{\rightarrow})^{-1}(\mathcal{M}')$ of $\mathcal{H}_{n,p}=\text{supp}^{-1}(p)$ is isomorphic to the Grassmannian $Gr(n,m)$. The geometric Langlands conjecture says that to $\mathcal{E}$ we can associate a $\mathcal{D}$-module $\mathcal{F}_\mathcal{E}$ on Bun$_m$, which is a Hecke eigensheaf of a given $\mathcal{E}$. For simplicity we wonsider $G=GL_1(\mathbb{C})$ or its compactification $G_c=U(1)$. Then $\text{Bun}_1$ is the Picard variety 
\begin{align}
\begin{aligned}
\text{Pic}&=\bigsqcup_{d=0}\text{Pic}_d\\
\text{Pic}_d&=\left\{\mathcal{L}\in\text{Pic}:d=\int_{T^2_{BZ}} c_1(\mathcal{L}) \right\}.
\end{aligned}
\end{align}
A rank 1 local system $\text{Loc}_1$ is a pair $\mathcal{E}=(\mathcal{L},\nabla)$ of a holomorphic line bundle and a flat connection. There is a natural map sending $\mathcal{E}$ to $\mathcal{\mathcal{L}}\in\text{Pic}_0$. Laumon \cite{Laumon} and Rothstein \cite{Rothstein1996} established the geometric Langlands correspondence by applying the Fourier-Mukai transformation \cite{mukai1981}: $\text{Loc}_1$ is transfered to $\text{Pic}_0$. In this case $h^{\rightarrow*}(\mathcal{F})$ is the Hecke modification of a $\mathcal{D}$-module $\mathcal{F}$. The operation $h^{\rightarrow}:T^2_{BZ}\times \text{Pic}\to \text{Pic}$ maps a holomorphic line bundle $\mathcal{L}$ as $(p,\mathcal{L})\to \mathcal{L}'=\mathcal{L}(p)$, by which $c_1(\mathcal{L}')=c_1(\mathcal{L})+1$. One can consider a more general modification of $\mathcal{L}$ to $\mathcal{L}'$ at $N$-tuple of points $(x_i),i=1,\cdots, N$ so that $c(\mathcal{L}')=c(\mathcal{L})+N$. 

To continue where we left off \cite{IKEDA2018136} we consider the 2d IQHE, which can be explained by holomorphic line bundles on $T^2_{BZ}$. We remind the statement below.
\begin{claim}
The plateaus of the Hall conductance
are the Hecke eigensheaves and quantized Hall
conductance is due to the Hecke translation
acting on line bundles.
\end{claim}
The local system is precisely
given by the Berry connection \cite{doi:10.1098/rspa.1984.0023} and a representation
of $\pi_1(T^2_{B
Z})$ is given by the Berry phase
\begin{equation}
\gamma_n[C] =\oint_C dR\bra{n,R}\nabla\ket{n,R},    
\end{equation}
 where the state $\ket{n,R}$ is adiabatically translated along a closed path $C$. The plateaus are exactly formed by the
wave functions which localize around impurities
in the system, called the Anderson localization
\cite{PhysRev.109.1492}, and such localized wave functions do not
carry Hall conductance. Hence the associated
gauge connections are flat, which corresponds
to Hecke eigensheaves. On the other hand, the quantization $\sigma_{xy} = \frac{e^2}{
h}n~(n \in \mathbb{Z})$ of Hall
conductance is given by the first Chern number
$\int_
{T^2_{BZ}}c_1(\mathcal{L})$ of a line bundle $\mathcal{L}$, according to
the TKNN formula \cite{PhysRevLett.49.405}. Moreover
$\int_
{T^2_{BZ}}c_1(\mathcal{L})$ 
corresponds to the order of a pole (vortex)
in $T^2_{B
Z}$~\cite{kohmoto1985topological}. The operator $h^\rightarrow$ acts on $\mathcal{L}$ as
$c_1(h^\rightarrow(\mathcal{L})) = c_1(\mathcal{L}) + 1$. This is why the Hall conductance experience quantum jump.

For a generic Lie group $G$, its Langlands dual
group $^LG$ is uniquely determined and the geometric Langlands conjecture expects that for
a $^LG$-local system $\mathcal{E}$ on $T^2_{BZ}$, there exists a corresponding Hecke eigensheaf FE on $\text{Bun}_G$.Note if $G = GL_m$, then the dual is isomorphic to $^LG = GL_m$. The geometric Langlands for $G = GL_m$ is proved in \cite{gaitsgory2004vanishing, 2000math.....12255F}. The IQHE or type A topological insulators are generically classified
by the Grassmannian $Gr(n,m)$ \cite{PhysRevB.78.195125}, which is consistent with
the classification of the Hecke correspondence.

What the geometric Langlands correspondence expects is that for a given flat $GL_1$-bundle $\mathcal{E}$ on $X$, there exist a unique $\mathcal{D}$-module $\mathcal{F}_\mathcal{E}$ on $\text{Pic}(X)$ associated with the modification $h_x$. This correspondence is proven by P. Deligne~\cite{Deligne1973}. The general conjecture of the Langlands correspondence for a Lie group $G$ can be stated as follows.  We denote by $^LG$ the Langlands dual group of $G$. If $G=GL_1$, then its dual is isomorphic to $GL_1$. The set $\text{Loc}_{^LG}(X)$ of local systems is again identified with the set of conjugacy classes of representations $\rho:\pi_1(X)\to\hspace{-1mm}^LG$. And $\text{Pic}(X)$ is generalized to the moduli stack $\text{Bun}_G(X)$ of principle $G$-bundles on $X$. So the geometric Langlands correspondence implies that for a given flat $^LG$-bundle $\mathcal{E}$ on $X$, there is a unique $\mathcal{D}$-module $\mathcal{F}_{\mathcal{E}}$, called a Hecke eigensheaf, defined on $\text{Bun}_G(X)$ associated with the Hecke modification. 

The tight-binding Hamiltonian of the IQHE in fractional magnetic flux $\phi=a/b$ can be written by the quantum group $U_q(sl_2)$, where $q=e^{2\pi ia/b}$ with coprime integers $a,b$ \cite{PhysRevLett.72.1890}. The strong/weak duality $(\phi,U_q(sl_2))\leftrightarrow (1/\phi,U_{{}^Lq}(sl_2))$ can explain the fractal energy spectra, called the Hofstadter butterfly~\cite{1976PhRvB..14.2239H}. Here $U_{{}^Lq}(sl_2)$ is the Langlands dual quantum group. There is a relation $\nu_L=\phi\nu_B$ between the tight-binding band filling factor $\nu_B=t\phi+s$ with $s,t\in\mathbb{Z}$ and the Landau level filling factor $\nu_L=t(1/\phi)+s$. Hence The strong/weak duality is $(s,t)\leftrightarrow (t,s)$ \cite{PhysRevB.67.195336}. The Widom-Str\v{e}da formula $\sigma_{xy}=-\frac{e^2}{h}\frac{\partial \nu_B}{\partial \phi}$ gives $(\phi, \sigma_{xy}=-\frac{e^2}{h}t)$ and $(1/\phi, \sigma_{xy}=-\frac{e^2}{h}s)$. More precisely, the strong/weak duality is a duality between momentum space in the flux $\phi$ to the real space in $1/\phi$. In this sense, the duality is also based on a picture of the Fourier transformation. Alternative description can be found in Ising model and topological codes of quantum computation. This point will be revisited when discussing Ising model. 

\section{Fractional Quantum Hall Effect}
Let $G=SL_2(\mathbb{C}),{}^LG=PSL_2(\mathbb{C})$ and consider their Lie algebras $\mathfrak{g}\simeq {}^L\mathfrak{g}\simeq sl_2$. The geometric Langlands correspondence is related to the WZW (Wess-Zumino-Witten) models as follows~\cite{Frenkel:2005pa,Teschner:2010je}:
\begin{equation}\notag
\begin{CD}
{}^LG\text{-local system}@. \leftrightarrow @.\text{Hecke eigensheaf on Bun}_G\\
@AAA @. @AAA\\
\text{WZW}_{\hat{k}}(sl_2)@. \leftrightarrow @. \text{WZW}_{k}(sl_2)
\end{CD}
\end{equation}
The Liouville parameter\index{Liouville parameter} $b$ is related to WZW models\index{WZW model} \cite{WESS197195,WITTEN1983422} by 
\begin{equation}
\hat{k}+2=\frac{1}{k+2}=b^2. 
\end{equation}
Vafa relates $b^2$ to $\nu$ based on M-theory and the $G$ Chern-Simons theory \cite{Vafa:2015euh}. In our notation it is $b^2=1/\nu$. The Chern-Simons theory is symmetric under $b\to1/b$, which is the modular duality of the Liouville theory \cite{Faddeev2001} and $S$-duality of the $SL_2(\mathbb{R})$ Chern-Simons theory \cite{Terashima2011}. Back to the WZW models and consider vertex operators $V_\alpha(z)=e^{i\alpha\phi(z)}$ of a scaler field $\phi$ in the two-dimensional CFT (Conformal Field Theory). The Langlands duality is often referred to the electric/magnetic (or charge/vortex) duality \cite{Kapustin:2006pk}. Indeed physicists constructed Langlands dual groups of Lie groups in the context of electric/magnetic duality \cite{0034-4885-41-9-001}. We rephrase it as the correspondence between the "electric" vertex and "magnetic" vertex \cite{Frenkel:2005pa,Frenkel02}:
\begin{align}
\begin{aligned}\notag
\text{Electric vertex~}&\leftrightarrow \text{Magnetic vertex~}\\
V_{\alpha_+}(z)\overline{V}_{\alpha_+}(\overline{z})&\leftrightarrow V_{\alpha_-}(z)\overline{V}_{\alpha_-}(\overline{z})
\end{aligned}
\end{align}

If we defined $\alpha_+=\sqrt{p/q},~\alpha_-=-\sqrt{q/p}$ with coprimes $p,q$, the FQHE filling factor $\nu$ is identified with $\alpha_+=1/\sqrt{\nu}$. \if{This identification is possible by reading the vertex operator of the FQHF's boundary CFT (Tomonaga-Luttinger liquid \cite{doi:10.1143/ptp/5.4.544,doi:10.1063/1.1704046,MILLIKEN1996309}) with the central charge $c=1$.}\fi We may write $\nu=N_e/N_\phi$ as the ratio of electrons $N_e$ to that of magnetic fluxes $N_\phi$. The standard composite particle pictures associate anyon excitation modes with vortexes, which are accompanied with the statistical gauge connections \cite{doi:10.1142/S0217979292000037,doi:10.1142/S0217979296000337,PhysRevLett.62.86}. For example, composite boson fields $\Phi(z)~z=(z_1,\cdots,z_n)$ obeying the Shr\"odinger equation $H\Phi(z)=E\Phi(z)$ generates a $\mathcal{D}$-module. Picking up vorticity would be the Hecke transformation. In this way, the charge/vortex duality plays a fundamental role for the FQHE and gives us a clear analogue of the arguments by \cite{Kapustin:2006pk}.

Langlands program also sheds light on knot theory. \if{The Jones polynomials \index{Jones polynomial} endows a knot with an invariant. It was physically formulated in \cite{witten1989} by Chern-Simons theory and more recently relations with the Khovanov homology \cite{khovanov2000}, which is a categorification of Jones polynomials, and the geometric Langlands duality are suggested in \cite{Witten:2016qzs,WItten:2011pz}. }\fi We investigate it in terms of the FQHE. First of all, anyons with charge $\nu$ carry the fractional Hall conductance $\sigma_{xy}\propto1/\nu$ and exchanges of their positions in the 2d system generate the braid group \cite{MOORE1991362}. The Knizhnik-Zamolodchikov equation (KZ-equaiton)~\cite{KNIZHNIK198483}, which is the differential equation of the vacuum expectation value of primary fields in the WZW model, is the corresponding integrable system. Let $\text{Conf}_n$ be the configuration space of different $n$ points in $\mathbb{C}$:
\begin{equation}\notag
\text{Conf}_n=\{(z_1,\cdots,z_n)\in\mathbb{C}^n: z_i\neq z_j~ \forall i\neq j\}.
\end{equation} 
Let $\{e_i\}_{i=1}^3$ be the basis of $sl_2(\mathbb{C})$ and $V$ be a representation space of $sl_2(\mathbb{C})$. Define $\tau_{ij}\in V^{\otimes n}$ by 
\begin{equation}\notag
\tau_{ij}=\sum_{k=1}^31\otimes\cdots\otimes1\otimes e_k\otimes 1\otimes\cdots\otimes1\otimes e_k \otimes 1\otimes\cdots\otimes1,
\end{equation}
where $e_k$ are inserted into the $i$-th and the $j$-th positions. The KZ-equation is a differential equation of $W:\text{Conf}_n\to V^{\otimes n}$ 
\begin{equation}
dW=\frac{1}{\kappa}\omega W
\end{equation}
where $\kappa$ is a complex parameter and $\omega=\sum_{i<j}\tau_{ij} d\log(z_i-z_j)$ is a differential one-form on $\text{Conf}_n$. $\kappa$ is related to $q$-parameter of $U_q(sl_2)$ as $q=e^{2\pi i/\kappa}$. We can make contact with the WZW model by choosing $\kappa=k+2$. Taking paths $\gamma(t)=(z_1(t),\cdots,z_n(t))\in \text{Conf}_n$ for $t\in[0,1]$ such that $\gamma(0)=\gamma(1)$, we obtain the braid group\index{braid group} $B_n\simeq \pi_1(\text{Conf}_n/\mathfrak{S}_n)$, where $\mathfrak{S}_n$ is the permutation group of $n$ positions. Parallel translation of $W(\gamma(t))$ along $\gamma(t)$ gives a generic monodromy representation of $B_n$\begin{equation}
\rho_{\text{KZ}}:B_n\to \text{End}(V^n). 
\end{equation}
2d irreducible representations $T_i=\rho_\text{KZ}(\sigma_i)$ of the generators of $B_n=\langle \sigma_1,\cdots, \sigma_{n-1}\rangle$ naturally generate the Iwahori-Hecke algebra $H_n(q)$ with $q=e^{2\pi i/\kappa}$. Take an $n$-tuple $(V_1,\cdots,V_n)$ of $sl_2(\mathbb{C})$ irreducible representations. A set of matrices $H=(H_1,\cdots,H_n)$ of the extended KZ-equation, where $H_i=\sum_{j}\rho(\tau_{ij})d\log(z_i-z_j)$ and $\rho:V^{\otimes n} \to V_1\otimes\cdots\otimes V_n$, defines the Gaudin model and differential equations $H\Psi=E\Psi~(i=1,\cdots,n)$ gives a $\mathcal{D}$-module \cite{Frenkel:1995zp}.

Finally we see how our viewpoints are consistent with several renowned works. It is possible to construct Jones polynomials by the 2d irreducible representation $\rho_{V_2}:U_q(sl_2)\to \text{End}(V_2)$, which is conjugate to $\rho_\text{KZ}$ \cite{AIF_1987__37_4_139_0}. We write $G=SL_2(\mathbb{C})$ and $G_c=SU(2)$. Jones polynomials can be obtained by $G_c$ Chern-Simons action $\frac{k}{4\pi}CS(A)$ \cite{witten1989}. An analogue of the KZ-equation is obtained by $G$ Chern-Simons action $S(\mathcal{A})=\frac{k+s}{8\pi}SC(\mathcal{A})+\frac{k-s}{8\pi}SC(\overline{\mathcal{A}})$ \cite{witten1991}, where $k\in\mathbb{Z}$ and $s$ is either real or pure complex. The Euler-Lagrange equation of the action $S(\mathcal{A})$ asserts that $\mathcal{A}$ is flat and thereby the moduli space of flat $SL_2(\mathbb{C})$ connections is exactly the symplectic manifold with a family of hyper K\"{a}hler structures. Tuning parameters of its complex structures leads to the Langlands duality in the context of Kapustin-Witten \cite{Kapustin:2006pk,Witten:2016qzs}. Categorifixation of the $(U_q(sl_2), V_2)$ knot invariants (Jones polynomials) is called the Khovanov homology \cite{khovanov2000} and its Langlands duality is investigated in \cite{Witten:2016qzs,WItten:2011pz}. The FQHE $\sigma_{xy}=\frac{e^2}{h}\nu$ can be explained by the Chern-Simons theory with $G$ or $G_c$. The $q$-parameter of $U_q(sl_2)$ accommodates $\nu$ as $q=e^{2\pi i/\nu}$ and the Langlands duality of $U_q(sl_2)$ implies flipping $\nu\to1/\nu$. The statement below summaries our discussion above.
\begin{claim}
The FQHE duality $(\nu, \rho_{KZ}(B_n)) \leftrightarrow (1/\nu, ^L\rho_{KZ}(B_n))$ is the Langlands duality of representations
of braid groups.
\end{claim}

\section{Remarks on Duality in Statistical Physics, Quantum Computation and Computational Optimization Problems}
Our story can be also applied to statistical physics. Indeed, the correspondence of the partition functions at low and high temperature is widely known among those specialists. To see this we consider spin models on a lattice. To focus on the Fourier transform, which is the basic starting point in Langlands duality, let us proceed as follows. This story is independent of any particular way of taking a lattice. Let $S_i\in \mathbb{Z}/q\mathbb{Z}$ be spin at lattice point $i$.
We first assume that the Boltzmann factor $f(S_i,S_j)$ of the nearest neighbor pair $i,j$ is written as 
\begin{equation}
    f(S_i,S_j)=f(S_i-S_j)~\text{mod}~q.
\end{equation}
Then in general the partition function can be written as 
\begin{equation}
    Z=\sum_{S_i}\prod_{\langle i,j\rangle}f(S_i-S_j),
\end{equation}
where $\langle i,j\rangle$ stands for the nearest-neighbor sites $i,j$. The analogue of the correspondence between the $L$-functions is now the correspondence between the statistical partition functions. In the most ideal case, we generally expect that it can be written as follows up to a constant factor:
\begin{equation}\label{eq:dual}
Z(K)=Z(K^*),
\end{equation} 
where $K=J/T$ is the inverse temperature and $K^*$ is dual to $K$ in the sense explained later. We are mostly interested in square lattice models but the same estimation is essentially true to a generic case. The duality \eqref{eq:dual} of partition functions is given by the quality 
\begin{equation}
\sum_{S}\prod_{\langle ij\rangle}u(S_i-S_j)=q^{-1-N}\sum_{S'}\prod_{\langle \hat{i},\hat{j}\rangle}\lambda(S'_{\hat{i}}-S'_{\hat{j}}),
\end{equation} 
where $u$ is the Boltzmann factor $u(S_i-S_j)$, $\lambda$ is the Fourier transformation of $u$, $S'$ is the spin variable on the dual lattice and $\{\hat{i}\}$ denotes lattice points dual to the lattice denoted by $\{i\}$ \cite{doi:10.1063/1.522914}. Note that this duality is in parallel with the duality of a real lattice and its reciprocal lattice. The Brillouin Zone (BZ) which is identified with torus $T^2$ for 2d-QHE is the Wigner-Seitz cell of the reciprocal lattice. Such a correspondence reminds us the Fourier-Mukai transformation of homomorphic bundles in terms of the Langlands duality.  

Some tips to relate the Langlands program with statistical physics are proposed in \cite{Frenkel:1995zp}, which works on Gaudin models, an example of $O(n)$-vector model. $n=1$ is the Ising model, $n=2$ is the XY model, and $n=3$ is the Heisenberg model. In this section we consider the Ising model from a viewpoint of the Langlands program. 

Duality of Ising models are classical in the following two meanings: 1) the relevant phenomena have been widely studied and many of them have been known since old times. The classical Langlands duality shall be thought of as a Langlands duality in the realm of classical spin systems (as opposed to quantum spin systems such as the Heisenberg model). 2) Ising model involves classical phase transition at $T>0$. Besides physical interests, our purpose here is to explain this duality should be understood as classical Langlands duality. This is achieved when we consider a representation of the quantum group $\mathcal{U}_q(sl_2)$, where $q$ is essentially given by the Boltzmann weight. $|q|\neq1$ explains classicalness of the theory. This is in contrast to our previous discussions including quantum Hall effect.  

\begin{figure}[H]
\centering
\includegraphics[width=8cm]{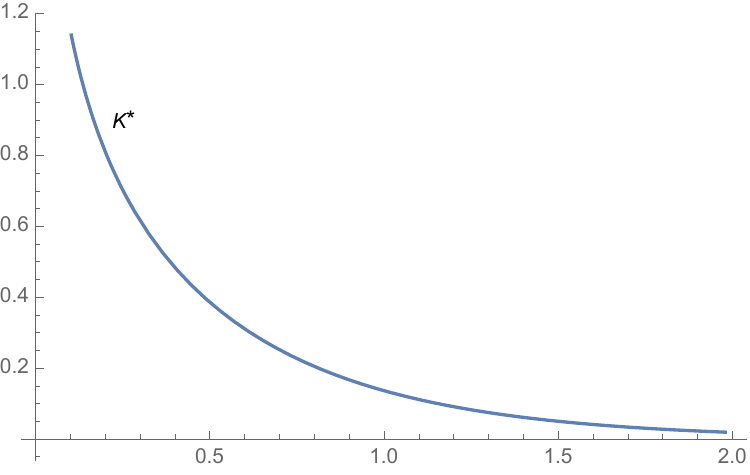}
\caption{The plot of $K^*$ against $K$.}
\label{fig:K}
\end{figure}

The partition function of the Ising model is the $q=2$ case. More precisely, in this case the duality \eqref{eq:dual} of partition functions is written as 
 \begin{equation}
 \frac{Z(K)}{2^N(\cosh K)^{2N}}=\frac{Z(K^*)}{2e^{NK^*}}, 
 \end{equation}
 where $K=J/T$, $N$ is the number of lattice points and $K^*$ is defined by 
 \begin{equation}
K^*=-\frac{1}{2}\log(\tanh K). 
 \end{equation}
 The relation between $K,K^*$ is not exactly the same as that between electric charge $e$ and magnetic charge $g$ which is $eg=1$, but it is very close as shown in Fig. \ref{fig:K}. Moreover taking logarithm expression
 \begin{equation}
 \frac{1}{N}\log Z(K)=\frac{1}{N}\log Z(K^*)+\text{(function without singularity)}
 \end{equation}
we can recover the duality formula $Z(K)=Z(K^*)$ up to the leading terms.  This formula claims the correspondence with high temperature regime $K$ and low temperature regime $K^*$. Here the non-singular part is negligible since it dose not contribute to phase transition. In addition the Fourier transformation of the Boltzmann factor\index{Boltzmann factor} clearly implies the correspondence $K\leftrightarrow K^*$ between low temperature and high temperature, that is, $u(1)/u(0)=e^{-2K}$ and $\lambda(1)/\lambda(0)=e^{-2K^*}$.  
 
The duality of Ising model is sometimes called the Kramers-Wannier duality~\cite{PhysRev.60.252}, by which the transition temperature $K=K^*=K_c$ can be derived. We understand that the Fourier transformation of the Ising model describes this duality of the Ising models. It is closely related to the Langlands program that the duality is derived by the Fourier transform. This Fourier transform not only relates the Boltzmann factors of the two models, but also maps the theory on the lattice to the theory on the dual lattice. Another way to see the duality is to consider the renormalization group. If we recall that the Langlands duality in the case of $GL_1$ on the torus is obtained by the Fourier-Mukai transform, there is a natural connection between the Kramers-Wannier duality and Langlands duality. 

The argument is almost in parallel with the duality of the Kondo effect~\cite{doi:10.1143/PTP.32.37}, which is the first known example of asymptotic freedom in physics associated with the interaction between the localized moment of magnetic impurities and conduction electrons becoming too strong to accommodate perturbations at low temperatures and low energies. The most established physical theory of asymptotic degrees of freedom is the theory of quark confinement, which is often described by the duality of 't Hooft operator and the Wilson operator. These two operators are exactly what Kapustin and Witten used to explain Langlands duality~\cite{Kapustin:2006hi,Kapustin:2006pk}. From this perspective, the localized moment of magnetic impurities in the Kondo effect is probed by the 't Hooft operator. In fact, some aspects of the Kondo effect can be seen in  QCD~\cite{PhysRevD.99.014040}. 

Now let us consider the duality of a quantum spin system. For simplicity  we address the Ising model
\begin{equation}\label{eq:Ising}
    H=-J\sum_i \sigma^z_i\sigma^z_{i+1}-h\sum_i \sigma^z_i
\end{equation}
The duality transformation is 
\begin{align}
\begin{aligned}
\tilde{\sigma}^z_i&=\sigma^x_1\sigma^x_2\cdots\sigma^x_i\\
\tilde{\sigma}^x_i&=\sigma^z_i\sigma^z_{i+1}\\
\tilde{\sigma}^y_i&=-i\tilde{\sigma}^z_i\tilde{\sigma}^x_i,
\end{aligned}
\end{align} 
 which obeys the same commutation relation of Pauli operators 
 \begin{equation}
     [\tilde{\sigma}^z_i,\tilde{\sigma}^x_i]=2i\tilde{\sigma}^y_i. 
 \end{equation}
 Using those operators one can find the Ising Hamiltonian \eqref{eq:Ising} is mapped to  
 \begin{equation}
     H=-h\sum_i \tilde{\sigma}^z_i\tilde{\sigma}^z_{i+1}-J\sum_i \tilde{\sigma}^z_i. 
 \end{equation}
Therefore the non-trivial duality $h/J\leftrightarrow J/h$ is satisfied in the exact sense. In fact, the point $h=J$ corresponds to the crittical temperature $K=K^*$ of the 2d classical Ising model. 
 
A study of duality has some practical meanings for quantum computation. In fact it is reported that the duality in the spin glass is highly related with quantum error correction \cite{2002JMP....43.4452D,2003AnPhy.303...31W}, in which topological properties of toric codes braiding with anyon-like excitation modes is a powerful tool for topological quantum computation (TQC). There are two approaches to TQC from condensed matter and quantum information. Non-Abelian anyons of the FQHE are practical candidates to realize the TQC theoretically \cite{freedman2003topological,Aharonov2009} and experimentally \cite{Kasahara,He294,PhysRevB.98.041105}. Algorithmic braiding (a CNOT-operation) around artificial defects on surface codes is also significant \cite{KITAEV20032}. Electric/magnetic duality is often referred to the Hadamard operation, which is a quantum version of the Fourier transformation. Once an Hadamard gate acts on a surface code on a graph, it is mapped to a surface code on the dual graph. This is the same as the duality of the 2d Ising model, which is realized by the Fourier transformation of Boltzmann weights assigned to vertexes \cite{PhysRev.60.252,doi:10.1063/1.522914}. Indeed such duality plays a fundamental role in the VDB (Van-D\"{u}r-Briegel) correspondence, which ensures universal computation using the 2d Ising model \cite{PhysRevLett.100.110501}. Adiabatic quantum computation \cite{2000quant.ph..1106F} or quantum annealing \cite{PhysRevE.58.5355} could be such candidates. Here a quantum annealer \cite{Johnson2011} is a practical quantum solver of combinatorial optimization problems. Quantum annealing uses the Hamiltonian dynamics of an initial quantum state to a sought-after final state that represents the lowest energy configuration of an encoded combinatorial optimization problem \cite{RevModPhys.90.015002}. It uses a transverse-Ising Hamiltonian $\sum_{ij}J_{ij}\sigma^z_i\sigma^z_j+\sum_i h_i\sigma^z_i$ by encoding a combinatorial problem into a sparsely connected graph expressing the hardware interactions \cite{venturelli2015quantum,2018arXiv181102524B,2017arXiv170801625N,2017arXiv171104889S,2016ISTSP..10.1053R,10.3389/fict.2017.00029,O'Malley18,ikeda2019application}.

\section{Discussion and Conclusion}
We have investigated the 2d QHE in terms of the Langlands program, including a geometric case, integrable systems and several algebras. The 2d IQHE is a $U(1)$ gauge theory and topological insulators with general gauge groups exhibits similar phenomena. The monopole or vortex like excitations are ubiquitous phenomena and some of them show intriguing properties. For example, the BKT (Berezinskii–Kosterlitz–Thouless) transition \cite{0022-3719-6-7-010,PhysRevLett.40.1727,PhysRevLett.47.1542} involves many vortexes at the critical point. Moreover it will be interesting to seek for some relations with a $K$-theoretic classification of topological insulators \cite{shiozaki2014topology,PhysRevB.78.195125}. Studies on topological matters  would endow the Langlads program with suggestions.

We have discussed the simplest case, and it should be natural to explore a generic case. One of the key ingredients is the topological terms (the Wess-Zumino-Witten (WZW) terms) associated with the non linear sigma models \cite{PhysRevB.78.195125}. And the geometric Langlands correspondence manifests power for investigating the WZW model \cite{Frenkel:2005pa}. This suggests that mathematical background of topological insulators would be much more fruitful than what it had been believed. 

Duality is universally crucial in physics. As we have discussed, electric/magnetic duality, strong/weak duality, and high/low duality can be explained by the Langlands duality. In this sense, the Langlands program is a grand unified theory of mathematics and physics. 

Before ending this paper, we would like to discuss some of the remaining themes for future research, with some recent developments. In this paper we have specifically discussed the geometric Langlands program for the case of a torus, but it can be extended to the theory of general Riemann surfaces. Riemann surfaces of genus two or more have negative curvature, and such Riemann surfaces can be constructed using hyperbolic planes. A band theory on the general Riemann surface has recently emerged, which can actually be verified using a quantum simulation technique called cQED (Circuit Quantum Electrodynamics)~\cite{doi:10.1126/sciadv.abe9170}. The theory of general Riemann surfaces is very well understood in algebraic geometry, and being able to verify it using actual physical property theory is a very useful development~\cite{2022arXiv220112689K,2021arXiv210710586I,Ikeda_2021,2022arXiv220106587A}. 

In the huge picture of the Langland Program, the duality dealt with this paper is only a small part of the picture. How are those dualities connected as pieces of the puzzle of the Langlands program in low energy physics, or in physical theory as a whole? Although cross-disciplinary research in algebraic geometry and low energy physics has just begun and there is not much prior research, the Langlands program could be a great guide for travelers without a chart.

\section*{Acknowledgement}
The author is supported by PIMS Postdoctoral Fellowship Award. The author thanks Steven Rayan for useful discussion.

\section*{References}
\bibliographystyle{utphys}
\bibliography{ref}
\end{document}